# Low dose CT reconstruction assisted by an image manifold prior

Guoyang Ma[c,1], Chenyang Shen[a,b,1], Xun Jia[a,b]

[a] Medical Artificial Intelligence and Automation (MAIA) Laboratory, University of Texas Southwestern Medical Center, Dallas, TX, 75235

[b] Department of Radiation Oncology, University of Texas Southwestern Medical Center, Dallas, TX, 75235

[c] Westwood High School, 12400 Mellow Meadow Drive, Austin, TX 78750

*Corresponding author: Xun.Jia@UTSouthwestern.edu

Abstract: X-ray Computed Tomography (CT) is an important tool in medical imaging to obtain a direct visualization of patient anatomy. However, the x-ray radiation exposure leads to the concern of lifetime cancer risk. Low-dose CT scan can reduce the radiation exposure to patient while the image quality is usually degraded due to the appearance of noise and artifacts. Numerous studies have been conducted to regularize CT image for better image quality. Yet, exploring the underlying manifold where real CT images residing on is still an open problem. In this paper, we propose a fully data-driven manifold learning approach by incorporating the emerging deep-learning technology. An encoder-decoder convolutional neural network has been established to map a CT image to the inherent low-dimensional manifold, as well as to restore the CT image from its corresponding manifold representation. A novel reconstruction algorithm assisted by the leant manifold prior has been developed to achieve high quality low-dose CT reconstruction. In order to demonstrate the effectiveness of the proposed framework, network training, testing, and comprehensive simulation study have been performed using patient abdomen CT images. The trained encoder-decoder CNN is capable of restoring high-quality CT images with average error of ~20 HU. Furthermore, the proposed manifold prior assisted reconstruction scheme achieves high-quality low-dose CT reconstruction, with average reconstruction error of < 30 HU, more than five times and two times lower than that of filtered back projection method and total-variation based iterative reconstruction method, respectively.

Keywords: Low-dose CT, Iterative reconstruction, Artificial intelligence, Deep neural network, Manifold learning

---

[1] The first two authors contributed equally.

## 1. INTRODUCTION

Ever since its introduction in the 1970s, x-ray Computed Tomography (CT) has become an important tool in medical imaging to obtain a direct visualization of patient anatomy (Kak and Slaney, 1988). Due to the use of x-ray, CT scans result in high radiation exposure to the patient, which leads to the concerns of lifetime risk of cancer (Hall and Brenner, 2008; de Gonzalez *et al.*, 2009; Smith-Bindman *et al.*, 2009). While the use of CT has demonstrated tremendous clinical values, the drawback of radiation exposure has counteracted its advantages, particularly for pediatric patients, who are more sensitive to radiation and have a longer life expectancy than adults (Brenner *et al.*, 2001; Brody *et al.*, 2007; Chodick *et al.*, 2007). Therefore, it is a central topic to develop effective approaches to reduce x-ray exposure while maintaining clinically acceptable image quality.

A simple way to lower the x-ray exposure is to reduce the mAs level in a CT scan. However, at the same time, this approach also reduces the number of x-ray photons reaching the detector, which increases the quantum noise of the x-ray projection data (sinogram). If the reconstruction algorithm remains unchanged, the quality of the reconstructed CT images would be deteriorated by the noise-contaminated projection data, yielding amplified noise and streak artifacts (Kak and Slaney, 1988). Over the years, numerous research efforts have been devoted to reconstructing CT images from noisy sinograms acquired at low-dose settings. Projection image domain approaches attempt to estimate the true sinogram from the contaminated measurements by applying certain image filters (Sauer and Liu, 1991; Demirkaya, 2001; Wang *et al.*, 2005) or by solving an optimization problem that effectively discriminate noise and the true sinogram structure (Tianfang *et al.*, 2004; La Rivière, 2005). Another group of methods tries to address the noise issue in the reconstruction process. While certain image filters can be directly applied to the reconstructed CT images for noise reduction purpose (Rust *et al.*, 2002), optimization based approaches are usually more effective, which reconstruct the CT image by solving an optimization problem using the noisy projection data subject to the condition of certain image properties. Wang et. al incorporated an noise model in the optimization problem and used a bilateral filter to constrain the solution image (Wang *et al.*, 2006). After the boom of compressed sensing (Donoho, 2006), image regularization approaches in this area have been extensively studied and applied to the CT reconstruction problem. In particular, Total Variation (TV) methods (Sidky *et al.*, 2006; Song *et al.*, 2007; Chen *et al.*, 2008; Sidky and Pan, 2008; Jia *et al.*, 2010b), edge preserving TV (Tian *et al.*, 2011b), tight wavelet frame (Jia *et al.*, 2011a), and non-local means (Jia *et al.*, 2010a; Tian *et al.*, 2011a) have all presented their tremendous power in the CT reconstruction context.

A key issue behind the success of these image domain approaches is that, the procedures are designed to effectively constrain the solution image quality to differentiate true CT image structures from contaminations, e.g. noise. For instance, the TV approach essentially favors those images that are piece-wise constant. The more effective these constraints are, in terms of describing the solution image properties, the more expected it is that the application of the constraint would be more capable of reconstructing high quality CT images under deteriorated data. Nonetheless, it is apparent that existing constraint terms are most designed to describe the desired image properties using certain mathematical expressions. Take the TV approach as an example, it employs the L1-norm of the image gradient to enforce a piece-wise constant image. While being effective to a certain extent, this may not necessarily be the best way to constrain the solution, as the mathematical expression only approximately describes the desired image property, but may not be able to fully capture the properties of real CT images.

Mathematically speaking, the true CT images reside on a low-dimensional manifold structure of the high dimensional linear space formed by image pixel values. The success of the aforementioned image domain processing approaches essentially makes assumptions about the structure of the low-dimensional subspace, describe it in a certain mathematical term, and uses that to constrain the solution image. The effectiveness of these methods lies in the accurate approximation of the true manifold. Recently, the burst of deep learning (LeCun *et al.*, 2015) and data science has shed a new light on this area. Deep neural networks are known to be extremely effective to accurately capture the underlying manifold structure by studying data points in the high-dimensional space, although the exact mathematical reason is yet to be fully understood. With the numerous available patient CT images accumulated over the years, it is potentially possible to derive the manifold of the real patient CT images using advanced deep neural networks. The derived manifold can be then employed in a CT reconstruction problem as a constraint to regulate the solution image quality and remove contaminations.

With this idea in mind, we propose a CT reconstruction approach that employs the learnt patient CT manifold as an image prior in this paper. The rest of this paper is organized as follows. Section 2 describes the methods, including the overall idea, manifold learning, iterative CT reconstruction and the experiments. The results of the manifold learning and iterative CT reconstruction are shown in Section 3. Section 4 will discuss a few issues such as the relations to other studies and the limitations and future work. Conclusions are drawn in Section 5.

## 2. METHODS

### 2.1 Overall idea

The main idea of the proposed framework is to regularize the low-dose CT reconstruction problem with a learned manifold of high-quality CT images. Low-dose CT reconstruction based on acquired sinograms is usually contaminated with noise, which degrades the image quality. We propose to constrain the iterative reconstruction process, so that the reconstructed CT will reside on the manifold, and at the same time maximally agree with the measured sinograms.

As such, we propose to first learn the manifold of true patient CT images using a deep neural network from a vast amount of real CT images. Once the manifold is learnt, it will serve as prior knowledge in the CT reconstruction process. The complete workflow of the iterative reconstruction is displayed in Fig. 1(a), in which $f$ and $g$ are the reconstructed image and measured sinogram, respectively. More specifically, a reconstruction algorithm only considering data fidelity with the projection data is first performed at each step to obtain a CT image based on a measured sinogram. Note that this reconstructed image is not necessary on the manifold, as the sinogram does not have the knowledge of the manifold. Hence, we seek for a similar CT image on the manifold prior and combine it with the reconstructed one via weighted summation. The combined results will then be fed into the next iteration step as the initial guess of reconstruction algorithm. This process is repeated, until a convergence condition is met. With a general manifold of high-quality CT images, we expect that image quality will be improved step-by-step and finally, an image retaining data fidelity with measurement and high image quality can be generated. This iterative process is graphically illustrated in Fig. 1(b). Note that the quality of the final image obtained using the aforementioned approach heavily relies on the generality and accuracy of the manifold. Therefore, learning the high-quality CT manifold is of central importance to the success of the proposed framework.

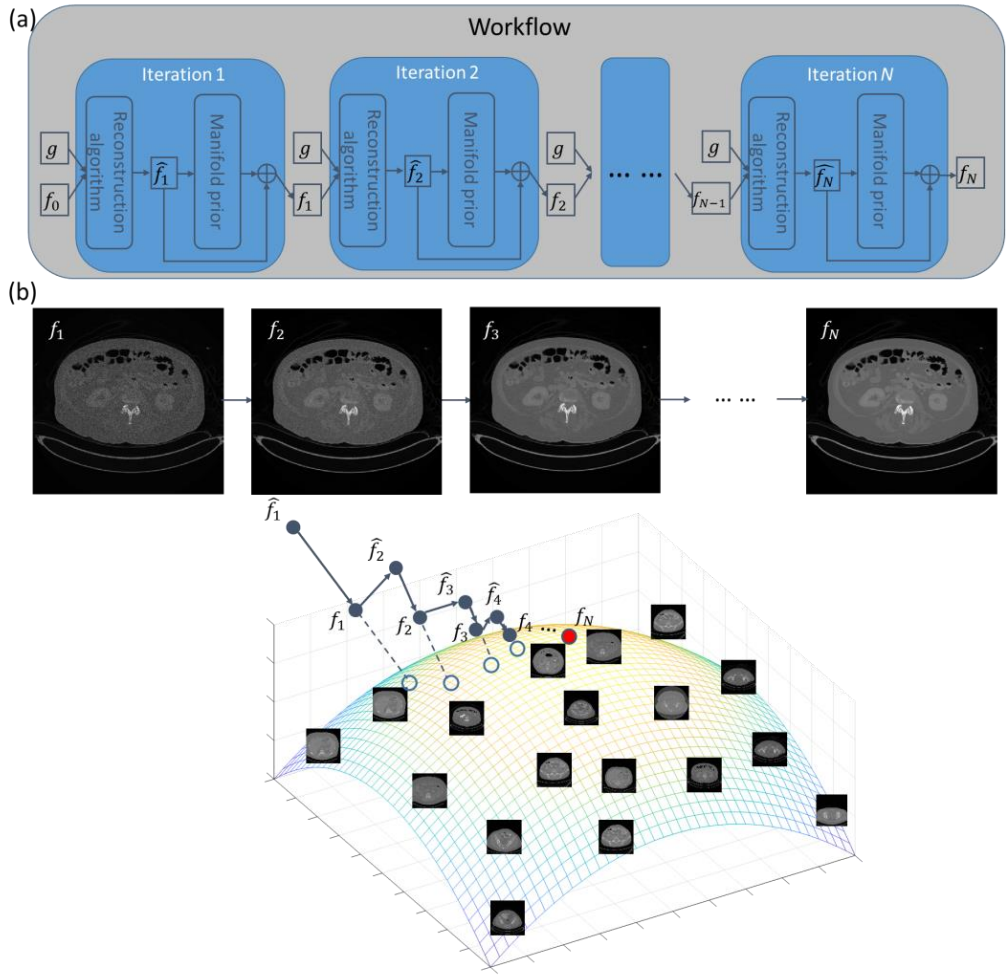

**Figure 1.** Main idea of the proposed reconstruction assisted by the manifold prior. (a). Workflow of the reconstruction framework. (b). A graphical illustration of reconstruction process. At each iteration, the reconstructed CT based on measurements and its projection on manifold are combined via weighted summation. It is then feed into reconstruction as initial guess. The algorithm converges to a CT image on manifold that maximally agrees with measured sonograms. $f$ and $g$ are the reconstructed image and measured sinogram, respectively.

*2.2 Manifold learning*

To incorporate the manifold prior into the reconstruction process as illustrated in Fig. 1, there are two essential tasks to be accomplished, i.e. to learn the mapping function of a CT image to the underlying manifold, and to restore the CT image from its manifold representation. Manifold learning has been studied intensively in the context of dimension reduction. Classical manifold (Roweis and Saul, 2000; He and Niyogi, 2004; He *et al.*, 2005) learning methods seek simple mapping functions from the original data space to some low-dimensional manifolds preserving local data structures. These methods usually suffer from inaccurate characterization of the manifold due to the limited complexity of the mapping function. Moreover, most of them are developed specifically for clustering or classification purposes, while restoring original data based on manifold representation is not possible.

Deep learning based methods have recently been adopted to address the limited capability of mapping function (Brosch *et al.*, 2013; Zhu *et al.*, 2016), which shed some light on manifold learning from a brand-new angle. In this paper, to achieve an accurate mapping of CT images to the underlying manifold, as well as high-quality image restoration from the manifold representation to the CT image simultaneously, we propose to incorporate an encoder-decoder deep convolutional neural network (CNN), because of its great ability in image feature extraction and representation. As shown in Fig. 2, the encoder network consists of seven convolutional blocks. Each block contains two convolution layers, each followed by a batch normalization layer (Ioffe and Szegedy, 2015) and a ReLu (Rectified Linear Unit) layer (Nair and Hinton, 2010) as nonlinear activation. A max-pooling layer is then employed at the end of each block to reduce the dimension of representation. This encoder network aims at mapping the original CT image to a manifold with a low-dimension representation compared to the high-dimension space of the CT image. The decoder network, designed as the inversion of the encoder, is constructed with a purely symmetric structure. Therefore, the decoder network also contains seven blocks, each of which starts with an upsampling network to increase the dimension, followed by two pairs of convolution layers and ReLu layers. In contrast to the encoder network, the decoder network is designed to restore the CT image from its low-dimensional representation on the manifold. The encoder-decoder CNN is trained simultaneously by enforcing an agreement between the output and input high-quality real CT images.

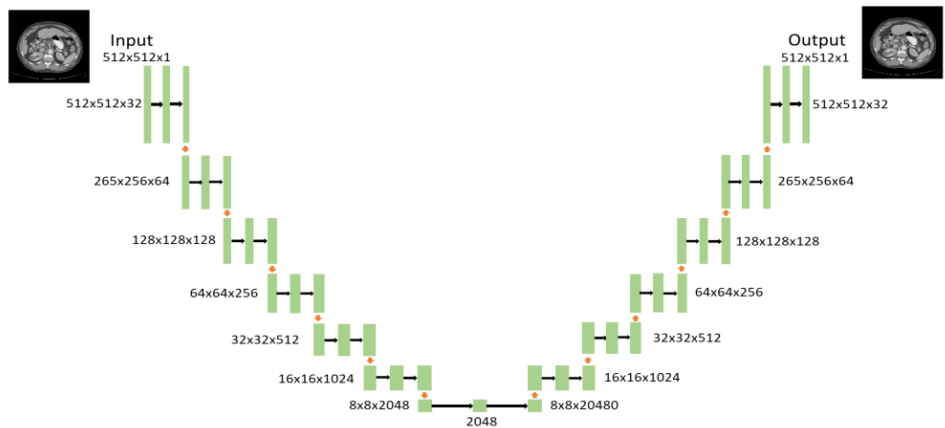

**Figure 2.** The encoder-decoder CNN for manifold learning.

More precisely, let $C(\cdot,\theta) = D(E(\cdot,\theta_E),\theta_D)$ denote the encoder-decoder CNN with $E(\cdot,\theta_E)$ and $D(\cdot,\theta_D)$ corresponding to the encoder network and the decoder network, respectively. The entire network $C(\cdot,\theta)$ is parameterized by $\theta = \{\theta_E, \theta_D\}$, which denotes a set of unknow parameters to be determined by training the network. Our goal is to search for a parameter set $\theta^*$, such that $f = C(f,\theta^*)$ for any high-quality CT images $f \in R^{n \times n}$, where $n$ is the number of pixels in each dimension of the CT images. We formulate the network training task as an optimization problem to minimize the discrepancy, i.e. $\theta^* = argmin_\theta L(f, C(f,\theta))$, where $L$ is the loss function computing the distance between the network output and $f$. In this study, square of L2-difference is used as the loss function for simplicity. The training can then be performed by simply executing the following iterative scheme

$$\theta_{i+1} = \theta_i - \delta \frac{\partial L}{\partial \theta}, \tag{1}$$

where $\delta$ indicates the learning rate and $i$ is the index of iteration. $\frac{\partial L}{\partial \theta}$ can be computed via standard back-propagation in a typical network training process. In addition, a large number of CT images are necessary for training to ensure the generality of manifold representation. As in many other studies, we utilize the stochastic gradient descent strategy to compute the gradient because of its high efficiency and low memory demands. Updating $\theta$ following Eq. (1) until its convergence, an encoder-decoder CNN $C(\cdot,\theta^*) = D(E(\cdot,\theta_E^*),\theta_D^*)$ can be established. Once this is achieved, it is possible to compute the low-dimensional manifold representation of any CT image as $z = E(f,\theta_E^*)$, and to restore the CT image as $f = D(z,\theta_D^*)$ from its manifold representation.

Note that we only require the learned manifold to preserve as much information as possible for a high-quality restoration without any manually defined constraint. The main reason is that the true structure of manifold is so far unknown, and hence we derive the manifold using a purely data-driven approach. Moreover, the dimension of the manifold is much lower compared to that of the CT images. As such, the proposed encoder network retains the ability to remove high-dimensional contaminations in image domain such as noise signals and hence benefit low-dose CT reconstruction. One potential concern is that a noisy image as input of the encoder network may bias the manifold representation away from that of the corresponding clean image. To address this concern, in next section, we develop a novel iterative algorithm to tackle the desired manifold representation step by step, and thereby facilitating a high-quality reconstruction for low-dose CT.

*2.3 Iterative CT reconstruction assisted by a manifold prior*

As shown in Fig. 2(b), the reconstruction of low-dose CT image is usually contaminated by noise and hence it may not reside on the manifold of high-quality CT. Due to the noise signal, a direct mapping using the proposed encoder network may result in a biased representation on the manifold and thereby an incorrect CT image will be restored from the manifold via the decoder network. To tackle this issue, we propose an iterative reconstruction algorithm to obtain a high-quality CT image that best matches with the acquired sinogram.

More specifically, the proposed algorithm is derived to solve the following least-square reconstruction problem with the manifold constraint implemented via the proposed encoder-decoder CNN:

$$f^* = \underset{f=C(f,\theta^*)}{argmin} \|Pf - g\|_2^2. \tag{2}$$

Here $P$ represents the system matrix. The proposed algorithm contains two essential steps at each iteration. For each iteration $i$ $(i > 0)$, we solve the least-square problem without considering the manifold constraint. The conjugate gradient (CG) method is applied to obtain an intermediate CT image $\hat{f}_i$ by taking $f_{i-1}$, the resulting image from the previous step, as the initial guess. This step is essentially to respect the data fidelity between the reconstructed image and the measured sinogram $g$. After that, $\hat{f}_i$ is mapped onto the manifold. A CT image with improved quality is computed as

$$f_i = \frac{\hat{f}_i + \beta C(\hat{f}_i, \theta^*)}{1+\beta}. \tag{3}$$

The main idea of this step is to combine $\hat{f}_i$ with the CT image restored based on the manifold representation of $\hat{f}_i$, while $\beta$ is a relaxation parameter to balance their contributions. Repeating the two aforementioned steps until the algorithm converges, a high-quality reconstruction $f^*$ residing on manifold with data fidelity maximally satisfied is expected. The detailed algorithm is summarized in Algorithm 1.

---

**Algorithm 1.** The proposed reconstruction algorithm

> **Input:** $P, g, f_0, \beta$ and tolerance $\sigma$
> **Output:** $f^*$
> **Procedure:**
> 1. Set $i = 0$;
> 2. Compute $\hat{f}_i = \text{CG}(P, f_{i-1}, g)$;
> 3. Compute $f_i = \frac{\hat{f}_i + \beta C(\hat{f}_i, \theta^*)}{1+\beta}$;
> 4. If $\frac{\|f_i - f_{i-1}\|_2}{\|f_{i-1}\|_2} < \sigma$, set $f^* = f_i$;
>    otherwise, set $i = i + 1$, go to Step 2.

## 2.4 Experiments

In our experiments, 2200 patient abdomen CT slices have been collected from the cancer imaging archive (TCIA) (Clark *et al.*, 2013). Among them, 2000 randomly selected images are used to train the encoder-decoder CNN to generate a CT image manifold prior, while the rest are considered for testing purpose. Moreover, 10% of the training data are employed as validation data to monitor the quality of the network during the training step, based on which, hyperparameters are tuned, such as learning rate. In our study, 500 epochs of training have been performed with batch size of 10 and learning rate of $1 \times 10^{-4}$. The well-trained network is then saved and its performance in low-dose CT reconstruction has been evaluated using the testing data set.

All the CT images have a resolution of 512×512 pixels and a pixel size of 0.879×0.879 mm². In our data set, images with different resolution are resampled to this resolution via bilinear interpolation. Low-dose CT x-ray projection data is simulated using the testing data. More specifically, a line-shape x-ray detector is considered with 768 elements covering a 40 cm range. The source-to-isocenter distance is set as 100 cm and isocenter-to-detector distance is 50 cm. The projection matrix $P$ is computed using Siddon's ray-tracing algorithm (Siddon, 1985) for 120 equal-space projections covering an angle range of $360^o$, while the sinograms are obtained using $g = Pf^* + n$. $f^*$ indicates the ground truth CT image and $n$ is Gaussian noise with zero mean and 3% variance determined by $Pf^*$ as in (Wang *et al.*, 2008). Reconstruction assisted by the manifold prior is performed for the simulated low-dose CT using Algorithm 1. The performance of the proposed algorithm is evaluated by comparing with the clinical standard filtered back projection (FBP) method (Gordon *et al.*, 1975; Herman, 2009) and a representative iterative total-variation (TV) based reconstruction (Jia *et al.*, 2010b).

The proposed framework is implemented using Python with TensorFlow (Abadi et al., 2016) on a desktop workstation equipped with eight Intel Xeon 3.5 GHz CPU processors, 32 GB memory and two Nvidia Quadro M4000 GPU cards.

## 3. RESULTS

### *3.1 Manifold learning*

After the encoder-decoder CNN is fully trained, we apply the network to both training images and testing images to investigate the quality of the learned manifold. Note that our goal is to map a CT image onto a low-dimensional manifold and then restore the high quality CT image based on its manifold representation, while maximally preserving image information. Comparisons are made between the input CT image and its corresponding output from the proposed network. Fig. 3 and Fig. 4 are examples from training and testing data sets, respectively. We find that the output images from the network preserve most of the structure in original input CT images, which demonstrates the capability of the proposed network in manifold learning as well as image restoration from the manifold.

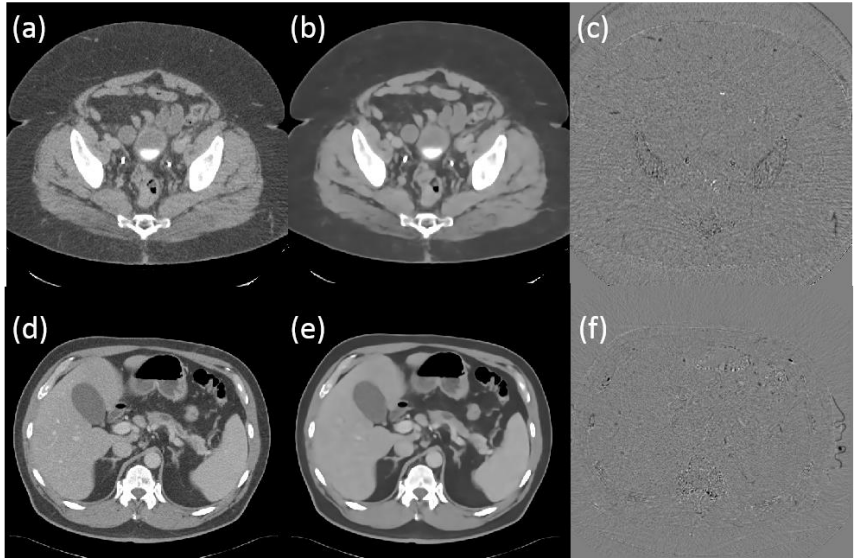

**Figure 3.** Performance of the proposed encoder-decoder CNN on two CT images from training data. (a) and (d) are input real CT images. (b) and (e) are their corresponding output images from the network. (c) and (f) give residual image between the input and output images. The CT images are displayed in a window [-200, 300] HU and residual images are displayed at [-150, 150] HU.

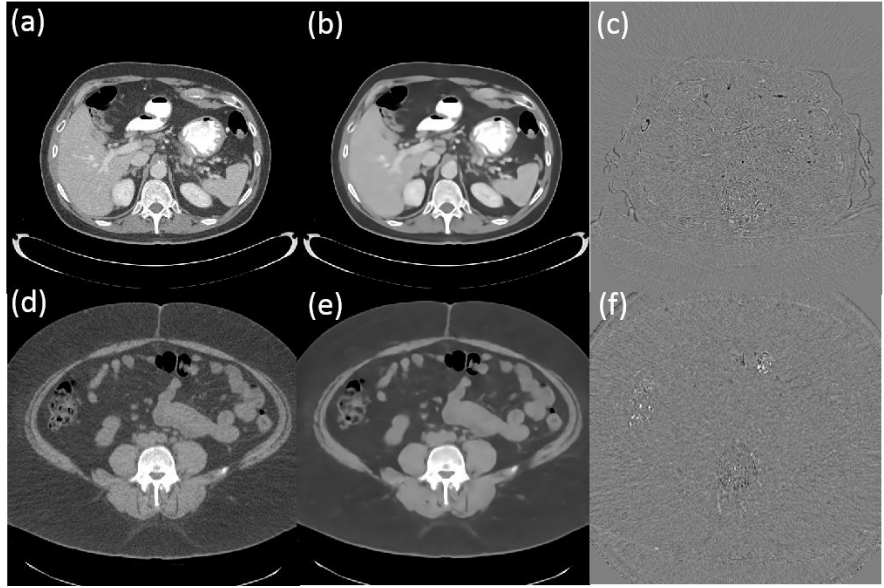

**Figure 4.** Performance of the proposed encoder-decoder CNN on two CT images from testing data. (a) and (d) are input real CT images. (b) and (e) are their corresponding output images from the network. (c) and (f) give residual image between the input and output images. The CT images are displayed in a window [-200, 300] HU and residual images are displayed at [-150, 150] HU.

In addition, Table. 1 summarizes detailed quantitative measures to evaluate the agreement between the input and out images. The average root-mean-square error (RMSE) on all the images are around 20 HU with average peak signal-to-noise ratio (PSNR) of >38. This again demonstrates the effectiveness of the proposed manifold learning method. Furthermore, encoder-decoder CNN performs well for not only the training images, but also for those testing images that have never been observed by the network in the training stage. According to Table. 1, there is no obvious degradation in the restoration accuracy from training data to testing data. This result validates the generality of the proposed network, i.e. the proposed network projects and restores general abdomen CT images accurately. With all these encouraging performance of the proposed manifold learning, we are ready to move on to test the proposed manifold assisted reconstruction in low-dose CT reconstruction.

**Table 1.** Quantitative results of manifold learning on both training and testing data.

|  | RMSE (HU) | PSNR |
|---|---|---|
| Training images | 20.2±7.6 | 38.26±2.26 |
| Testing images | 20.4±8.7 | 38.25±2.55 |
| Overall | 20.2 | 38.26 |

*3.2 Iterative CT reconstruction with manifold prior*

In this section, we report the effect of the encoder-decoder CNN as manifold prior in low-dose CT reconstruction (Algorithm 1). Fig. 5 and Fig. 6 show representative results of low-dose CT reconstruction on the training and testing data, respectively. It is easy to observe that the reconstruction quality using the proposed algorithm assisted by manifold prior significantly outperforms FBP and TV-based methods.

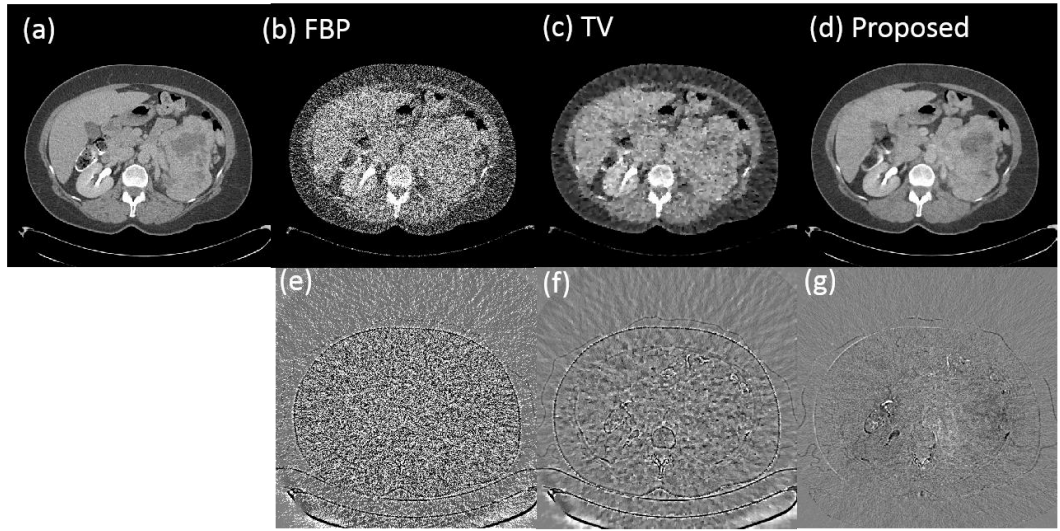

**Figure 5**. Comparison of reconstruction results on a training sample. (a) Ground truth (normal dose CT); (b) Reconstruction result of FBP method; (c) Reconstruction result of TV-based method; (d) Reconstruction result of proposed method; (e)-(f): residual image of FBP, TV, and the proposed method, respectively. The CT images are displayed in a window [-200, 300] HU and residual images are displayed at [-150, 150] HU.

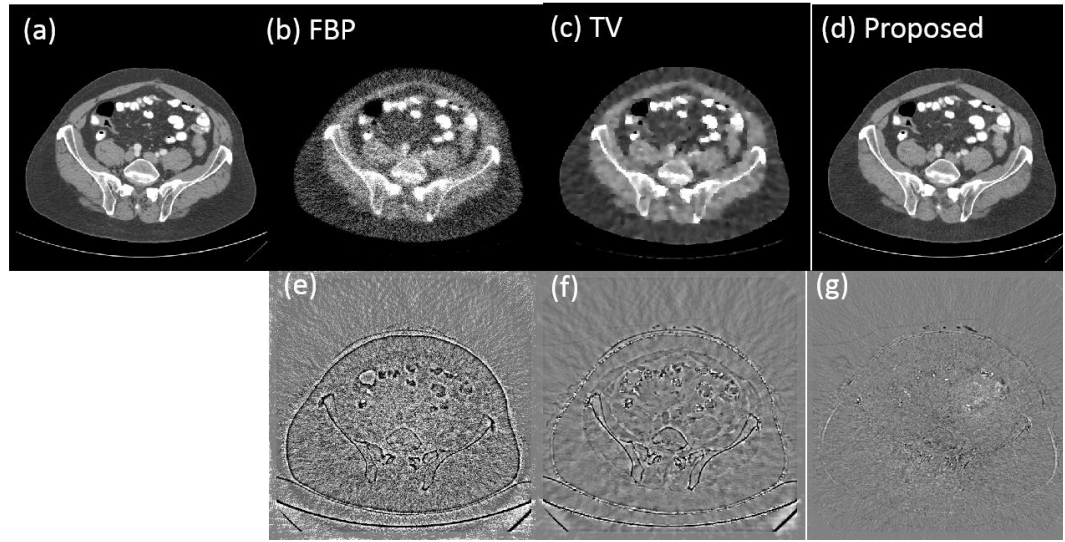

**Figure 6.** Comparison of reconstruction results on a testing sample. (a) Ground truth (normal dose CT); (b) Reconstruction result of FBP method; (c) Reconstruction result of TV-based method; (d) Reconstruction result of proposed method; (e)-(f): residual image of FBP, TV, and the proposed method, respectively. The CT images are displayed in a window [-200, 300] HU and residual images are displayed at [-150, 150] HU.

Quantitative evaluations of the low-dose CT reconstruction results using both training and testing images are summarized in Table 2. The advantage of the proposed method can be observed from both RMSE and PSNR. Compared to the FBP and TV-based reconstruction methods, the proposed algorithm at least obtains an improvement of ~9 in PSNR with more than 40 HU in reconstruction accuracy. In addition, we find that there is no significant difference between the results on training and testing data sets, showing the generality of the proposed method.

**Table 2.** Quantitative results of reconstruction on training and testing data.

|  | RMSE (HU) | | | PSNR | | |
| --- | --- | --- | --- | --- | --- | --- |
|  | FBP | TV | Proposed | FBP | TV | Proposed |
| Training | 177.3±20.4 | 122.3±14.6 | 38.8±22.8 | 21.56±4.79 | 23.70±1.32 | 33.40±3.35 |
| Testing | 175.2±17.5 | 115.7±17.2 | 35.8±16.8 | 21.71±2.04 | 24.25±1.99 | 34.49±1.91 |
| Overall | 177.1 | 121.7 | 38.5 | 21.57 | 23.75 | 33.50 |

## 4. DISCUSSION

### 4.1 Relation to other studies

The success of many iterative CT reconstruction approaches essentially incorporates some prior knowledge regarding image quality and uses that to regularize the solution image. For instance, the widely used TV based methods (Rudin *et al.*, 1992; Jia *et al.*, 2010b; Tian *et al.*, 2011b) assume high-quality CT images to be piece-wise constant in the image domain, while the tight-frame (TF) approaches (Jia *et al.*, 2011b; Dong *et al.*, 2013) rely on the hypothesis that images are sparse under wavelet transform. Such handcraft prior knowledge built based on of human observation usually incorporates some simple functions, e.g., gradient or wavelet transform. Although being very effective, they sometimes fail to robustly address the complicate structures of image data to a highly accurate level, as the mathematical description of features only approximate the true features, but not necessarily the most accurate.

Recently, deep learning has achieved tremendous success in many image oriented applications (LeCun *et al.*, 2015; Greenspan *et al.*, 2016; Han *et al.*, 2016; Chen *et al.*, 2018; Shen *et al.*, 2018). Deep learning-based CT reconstruction approaches (Han *et al.*, 2016; Chen *et al.*, 2017a; Chen *et al.*, 2017b) take advantage of powerful deep CNN architectures to represent CT images with superior quality and the advantage over handcrafted approach has been clearly demonstrated. Yet, study exploring the general low-dimensional manifold of CT images is still missing. Existing manifold learning algorithms (Roweis and Saul, 2000; He and Niyogi, 2004; He *et al.*, 2005) attempt to approximate the manifold structure incorporating manually defined constraints. Further study is still needed to evaluate whether these constraints truly hold on the manifold. In fact, a true manifold preserves all the useful information of the original CT image space and hence a restoration from manifold representation should be possible and its self-consistency between the original image needs to be established. Therefore, it is our main goal to facilitate mapping of both ways between the CT image space and the low-dimensional manifold by making use of the remarkable capability of CNN to approximate complex functions. The fully data-driven approach successfully establishes the desired two-way mappings, with its effectiveness demonstrated in preliminary experimental studies.

*4.2 Limitations and future work*

Despite the preliminary success of the proposed manifold prior based low-dose CT reconstruction, our framework has several limitations. First, although a realistic noise simulation (Wang *et al.*, 2008) method is utilized to generate low-dose CT projection data to test our approach, further investigation on real data is needed to further evaluate the effectiveness and robustness of the proposed algorithm. Second, the study learned the patient CT image manifold using only abdomen images, and it would be not surprising that the manifold would fail to constraint CT reconstruction in other body sites. One potential direction is to build body-site specific manifold and apply the corresponding one in each reconstruction problem. Meanwhile, it is also noted that the manifold for all the sites would form a universal one, which is expected to be applicable to the reconstruction problem of all the sites. While it is our future work to employ more CT images covering all the body sites to train this universal manifold, computational burden may become a concern.

Accurate and robust manifold prior has its value in the general context of CT reconstruction. Presumably CT images contaminated by any types of artifacts should not reside on the manifold of high-quality CT and hence their quality can be improved using the proposed framework. Future study will be performed to investigate the ability of using the manifold prior to handle the reconstruction in the presence of other types of projection contaminations, such as x-ray scatter, beam-hardening, and metal artifacts.

We have performed preliminary experimental study developed for general CT images, only using 2D abdomen CT slices. Manifold learning for 3D CT volumes is warranted as a future direction. Of course, substantial increase in computational load in both network training and reconstruction is expected. We will focus on developing GPU-based acceleration algorithms to boost the training and reconstruction efficiency.

## 5. CONCLUSION

In this study we developed a novel low-dose CT reconstruction framework assisted by a CT manifold prior. A deep encoder-decoder CNN has been constructed via a fully data-driven approach to effectively map a CT image to low-dimensional manifold, as well as to restore the CT image based on its manifold representation. A novel algorithm has also been designed and implemented to incorporate the manifold prior in low-dose CT reconstruction. We evaluated the performance of the proposed framework through a comprehensive simulation study in low-dose CT reconstruction. The proposed approach has been demonstrated to be capable of achieving superior low-dose CT reconstruction quality as compared to the clinical standard FBP algorithm and the typical TV-based iterative reconstruction algorithms.